# Deep Learning for automated phase segmentation in EBSD maps. A case study in Dual Phase steel microstructures.

T. Martinez Ostormujof [1,2], RRP. Purushottam Raj Purohit[1,2], S. Breumier[1,3], N. Gey[1,2], M. Salib[4], L. Germain[1,2], *

[1] Université de Lorraine, CNRS, Arts et Métiers ParisTech, LEM3, F-57000 Metz, France
[2] Université de Lorraine, Laboratory of Excellence on Design of Alloy Metals for low-mAss Structures ('LabEx DAMAS'), F-57073 Metz, France
[3] Institut de Recherche Technologique Matériaux, Métallurgie et Procédés, 4 rue Augustin Fresnel F-57078, Metz, France
[4] ArcelorMittal Maizieres, Research and Development, Voie Romaine, BP30320, F-57283 Maizieres-les-Metz, France

* E-mail address: lionel.germain@univ-lorraine.fr

Abstract

Electron Backscattering Diffraction (EBSD) provides important information to discriminate phase transformation products in steels. This task is conventionally performed by an expert, who carries a high degree of subjectivity and requires time and effort. In this paper, we question if Convolutional Neural Networks (CNNs) are able to extract meaningful features from EBSD-based data in order to automatically classify the present phases within a steel microstructure. The selected case of study is ferrite-martensite discrimination and U-Net has been selected as the network architecture to work with. Pixel-wise accuracies around ~95% have been obtained when inputting raw orientation data, while ~98% has been reached with orientation-derived parameters such as Kernel Average Misorientation (KAM) or pattern quality. Compared to other available approaches in the literature for phase discrimination, the models presented here provided higher accuracies in shorter times. These promising results open a possibility to work on more complex steel microstructures.

Graphical abstract

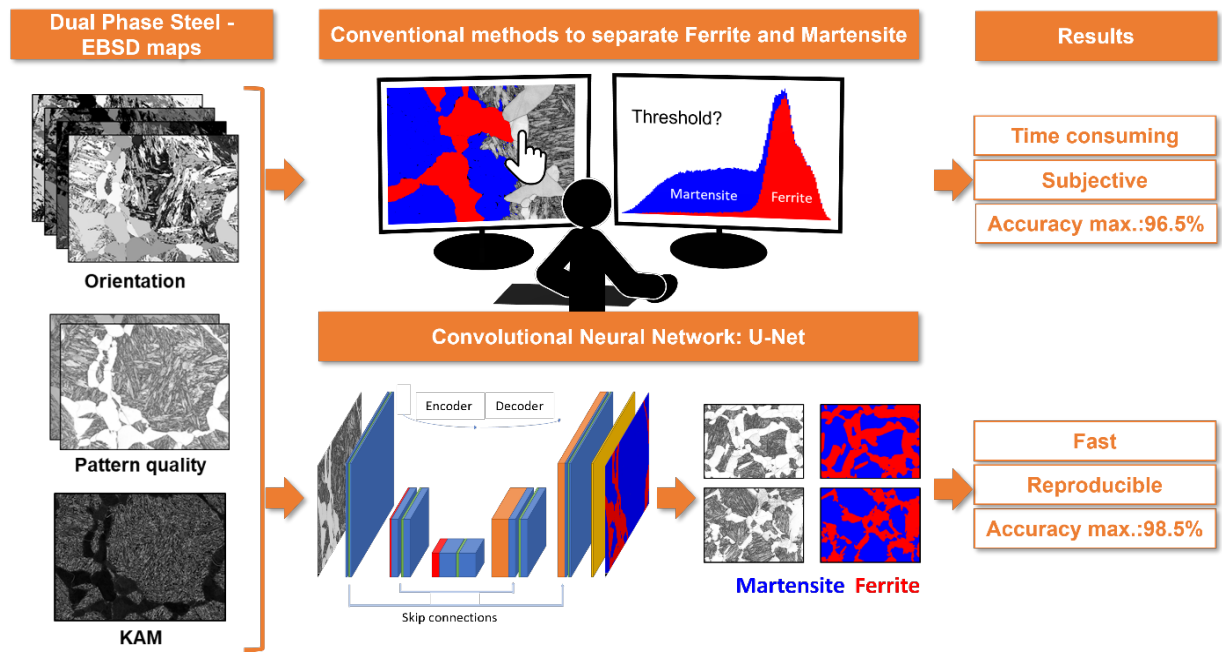

1. Introduction

The microstructural characterization of steels is key to optimize their processing route and ultimately better control their properties. It is of particular interest to identify, classify and quantify the different transformation products such as ferrite, pearlite, martensite and the different types of bainite inside the microstructure. Although Electron Backscattering Diffraction (EBSD) provides both microstructural and crystallographic information, it cannot discriminate those transformation products, being all indexed as body centered cubic (BCC). Nevertheless, the raw orientation data provided by EBSD intrinsically contains valuable information to distinguish the phases in steels: the presence of crystal defects, the existence and type of orientation relationship with the neighborhood, the spatial distribution of variants in packets/blocks, among others, are some important indicators [1]–[4]. Several advanced data processing approaches have been developed to extract these EBSD-based descriptors to semi-automatically discriminate steels' transformation products [5]–[15]. However, this task requires expertise, time and effort, and still may carry a certain degree of subjectivity.

In the search for alternatives, Artificial Intelligence (AI) based techniques are worth to be explored as they may potentially enable a quicker and effortless classification by unifying a criterion to reduce subjectivity. AI has already been applied to address the automatic characterization of steels, but mainly based on micrographs, and only to a lesser extent, on EBSD-based data.

In Machine Learning (ML) algorithms, a great effort must be dedicated to engineer and extract significant features or attributes of the phases in order to feed the algorithms. Both morphological and gray-level features extracted from optical and scanning electron micrographs have been proposed for steel's microstructural characterization [16]–[19]. Crystallographic features obtained by EBSD as variant pair density, Kernel Average Misorientation (KAM) and pattern quality have also been fed to ML algorithms to classify steel's microstructures [20], [21].

Deep Learning (DL) approaches emerged more recently and have the advantage to learn high-level features from raw input data avoiding the tedious feature extraction step. In particular, Convolutional Neural Networks (CNNs) [22] have been developed for image classification and segmentation since they can intrinsically capture the spatial dependencies of the data. Larmuseau et al. [23], [24] applied the ResNet CNN [25] and triplet networks on optical micrographs to classify a large variety of austenitic, martensitic and ferritic microstructures. However, as an image classification approach, it cannot distinguish different phases in the same micrograph. Ajioka et al. [26] implemented encoder-decoder CNNs as SegNet [27] and U-Net [28] to segment pixel-wise ferrite, martensite and grain boundaries (GBs) within optical micrographs. Although the presented results seem to overcome established state-of-the-art approaches, the presented microstructure appears to be rather simple and there is no tentative to generalize the model's performance into different testing cases. Azimi et al. [29] proposed Fully CNNs (FCNNs) as a method to segment SEM micrographs into martensite, tempered martensite, bainite, pearlite and ferrite. The method allowed accurate classification of ferrite (94% accuracy) but had more difficulties do distinguish other phases (< 80% accuracy). This highlights that microstructural data only may not be rich enough to carry out such a semantic segmentation task with a high accuracy rate. Kaufmann et al. applied CNNs to EBSD patterns (EBSPs) in order to identify their crystal symmetry [30], [31] and discriminate ferrite and martensite based on their subtle difference in lattice parameter variations [32]. Nevertheless, this strategy is applicable only when EBSPs have been saved (high storage capacity needed) and as in [26], no generalization to different testing examples is presented.

To the best of the authors' knowledge, no DL approach has been applied to raw orientation data to perform steels' phase classification. EBSD maps colored with the Inverse Pole Figure (IPF) key (RGB images), have been fed into CNNs in the past for other applications [33] but, since the color in the IPF is determined from the projection of a macroscopic direction in the crystal reference frame, it intrinsically reduces the orientation information. Thus, the main purpose of this contribution relies on applying CNNs to evaluate the ability of these networks to handle raw orientation data and compare its performance when trained with parameters derived from orientation data as pattern quality and KAM. A U-Net architecture was selected to perform supervised semantic segmentation on EBSD maps.

As a case of study, a Dual Phase (DP) microstructure formed by ferrite and martensite has been intentionally selected to be able to label as unambiguously as possible the EBSD maps and to reliably quantify the performance of the trained model. This approach is compared to other state-of-the-art methods [11], [34] and its performance is then evaluated on microstructures with higher complexity than those used for training. An upcoming paper will deal with the extension of this approach to segment also several type of bainites within multi-phase microstructures.

2. Experimental

2.1 Material

A model DP steel with composition (wt%) 0.17% C and 1.5% Mn has been selected. Three samples that underwent the heat treatment shown in Figure 1 were provided by ArcelorMittal Global R&D (Maizières-les-Metz, France). The treatment has been performed on a dilatometer Bahr. After an austenitization at 1000°C, the samples were hold at different temperatures to form proeutectoid ferrite and then quenched to room temperature to transform prior austenite grains (PAGs) into martensite.

The resulting microstructures consist in different ferrite-martensite fractions and morphologies (Figure 1). As expected, the fraction of martensite increases with the holding temperature. Whereas allotriomorph ferrite is observed at all temperatures, additional Widmanstätten ferrite is locally observed at T3=720°C. This ferrite nucleates on allotriomorph ferrite and grows inside the PAG with which it is in orientation relationship (OR) with a plate or needle morphology [35]. Upon quenching, martensite precipitates alongside these Widmanstätten plates and share nearly the same orientation.

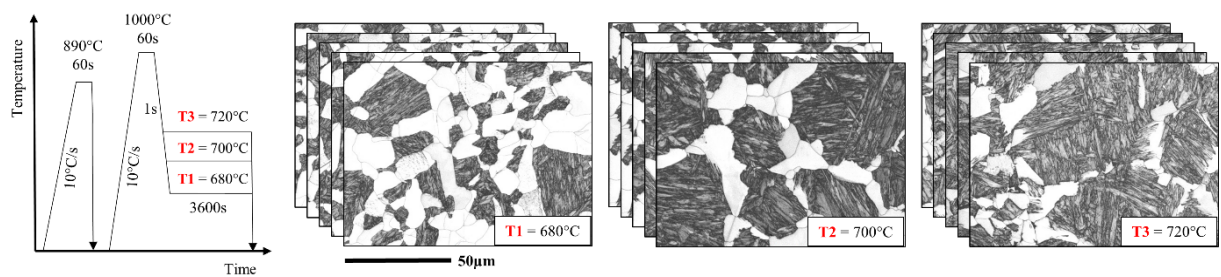

Figure 1. Heat treatments and corresponding microstructures (obtained from EBSD data).

2.2 EBSD acquisition

All 3 samples have been metallographically polished down to 1 µm diamond paste and a finishing step with colloidal silica (OP-U) was carried out. EBSD acquisition was performed on a ZEISS Auriga 40 FIB–SEM microscope equipped with the Oxford-instruments' CMOS Symmetry camera and AZtec acquisition software. A total of 5 maps per sample has been acquired at an accelerating voltage of 20 kV with an aperture size of 120 µm. The camera resolution was set to 156×128 pixels (i.e. 8×8 binning), the exposure time at 1ms and no frame averaging was used. The step size was set at 0.1 µm and the magnification at 1000×, resulting on a map size of 114.1×85.6 µm². The acquisition speed per point was ~970 Hz and the time needed to acquire an entire map was ~16 min. The resulting indexing rate was ~90%.

2.3 EBSD maps labeling

Labeling consists in assigning its corresponding ground truth phase, "Ferrite" or "Martensite", to each pixel of the map. Several indicators can be extracted from EBSD maps to assist phase discrimination (Figure 2).

The Band Contrast (BC) and Band Slope (BS) are two EBSD pattern quality indices directly available in AZtec (Figure 2a and b). The BC corresponds to the brightness of the bands and the BS evaluates the gradient of intensity at the edge of the bands [6]. These are expressed on a gray scale where black means the worst possible quality and white refers to the highest quality pattern. In our case, both are powerful parameters since martensite contains more dislocations, thus lower BC and BS values are expected [9], [11].

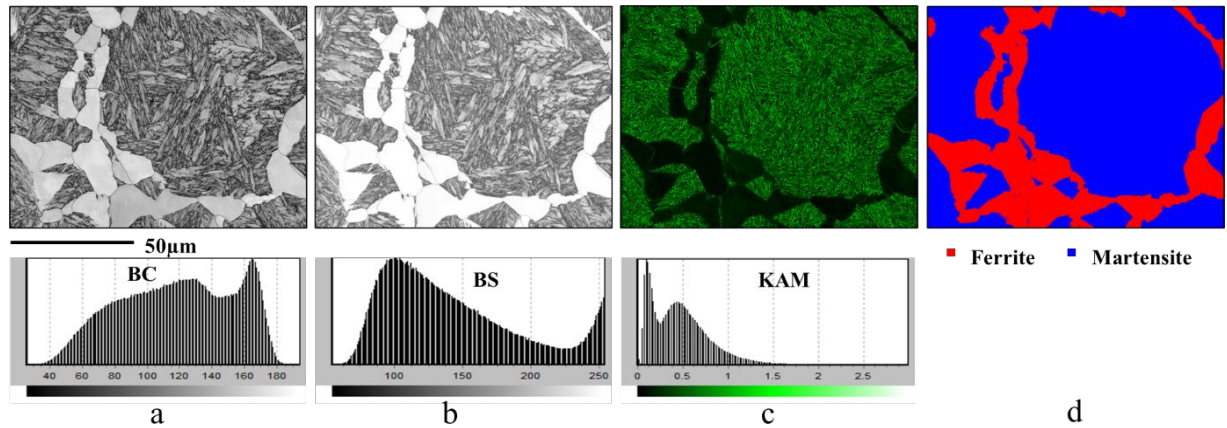

Figure 2. Specific EBSD indicators used to distinguish ferrite from martensite. Each indicator is plotted together with its histogram: a) BC, b) BS and c) KAM. d) Ground truth labeling.

The KAM also contains useful information to discriminate ferrite from martensite. It is computed as the average misorientation angle of the central pixel in a square kernel (here 3×3) relative to all other pixels in the kernel but excluding misorientations with an angle above a certain threshold (here 3°) (Figure 2c). The KAM has been recognized as a good qualitative estimator of the geometrically necessary dislocation density. Thus, higher KAM values are observed in martensite than in ferrite [36], [37]. This is confirmed on our data set: the average KAM for martensite is ~0.5° against ~0.2° for ferrite.

To label the 15 maps dataset a 3 step process has been carried out: 1) pre-labeling using Trainable Weka Segmentation (TWS) [34], 2) labeling refinement based on semi-automatic orientation data analysis and 3) non-indexed pixels corrections. 1) TWS has been used to train a classifier based on the BS images of 3 maps (one for each sample). The main idea of this tool is to manually label some regions that belong to ferrite and martensite which are then used to train a classification model (we have selected a Random Forest [38] classifier with all other settings left as default). This model has been applied to the entire dataset producing a pre-labeling where 0 and 1 were assigned to ferrite and martensite, respectively. 2) If the majority of pre-labeled pixels within a grain were ferrite, the grain was labeled as ferrite; martensite otherwise. The grains were detected at 2° closing boundaries down to 0° using the AlGrID algorithm [39]. This procedure was carried out to facilitate the identification of ferrite/martensite interphase boundaries that were weakly misoriented. However, most of the remaining errors concerned Widmanstätten ferrite/martensite interphase boundaries that were not separated during grain detection and manual corrections had to be performed, making Widmanstätten ferrite labeling somewhat subjective at the interphase boundary. 3) Non-indexed pixels were treated separately. By default, they were assigned to martensite since they were mainly located within martensite grains (lower quality patterns). However, an erode/dilate procedure was carried out to clean remaining errors at the ferrite/ferrite grain boundaries where certain non-indexed pixels could be wrongly assigned to martensite. This whole process has taken several hours to be done. An example of the final labeling is shown in Figure 2d.

3. Deep Learning strategy

3.1 Input EBSD data

Four models have been trained based on: BS, BC, KAM and raw orientation data as quaternions. Although orientations can be expressed in many ways, quaternions have been selected for several reasons: 1) their components are naturally in the [-1,1] range, 2) quaternions-based computations

(misorientation, averaging, etc.) involve only linear operations (products and additions), 3) they are very practical for high symmetry (cubic) computations [40] and 4) they are widely used by the EBSD community [41]–[44]. Since quaternions are described by 4 parameters, an adaptation of the U-Net architecture was required, as it normally works on RGB images (3 parameters).

## 3.2. EBSD data preprocessing

BS, BC and KAM values have been standardized (zero mean and unit standard deviation) since it has been demonstrated that it improves models training [45]. The orientation data was used "as acquired" i.e. no data cleaning was applied. The KAM has been calculated with a 3×3 pixels kernel size and a threshold angle of 3°. Non-indexed pixels were assigned with a zero KAM and with a quaternion whose components are all zero values. These zero values do not describe neither a KAM nor an orientation and highlight only the non-indexed points.

Quaternions were preprocessed so that all the orientations within the same grain were expressed with the same symmetry equivalent. In practice, grains are identified as a cluster of adjacent pixels disorientated one to the next by less than a threshold angle set by the user (here 3°) [46]. During this loop, we checked that all grain orientations were expressed with the same symmetry equivalent. If not, all the symmetry elements of the cubic group were iterated until this condition was met and the quaternion was changed accordingly. This process, unlike expressing all the quaternions in the fundamental space, guarantees the continuity of the orientation space inside the detected grains.

## 3.3 Dataset and data augmentation

To build any AI model data must be separated in 3 sets: training, validation and testing. The training set is used to optimize the model's parameters (weights of the convolutional filters) while the validation set is used to evaluate the model's performance during training. If the performance improves on the training set but remains constant on the validation set, the model is overfitting and training should be stopped. Once the final model is obtained, its performance must be evaluated on a testing set containing data never seen during training.

Data was split as follows: 2 EBSD maps from each sample were randomly selected for training and validation. This dataset contains a diversity of ferrite and martensite cases and sample preparation artifacts that were expected to be learnt in order to avoid their misclassification. The entire set of training-validation maps is formed by ~5.8 million pixels where ~2.7 million have been labeled as ferrite and ~3.1 million as martensite. Therefore, the training-validation set is balanced i.e. it contains a close number of ferrite and martensite pixels and the risk to bias the model towards one or the other phase is avoided. 9 maps (3 maps per sample) remained as testing set.

To increase the size of the training-validation set, data augmentation techniques were employed. New maps were created by performing rotations of 180° around x, y and z axes and 90° and 270° around the z axis of the 6 maps. The rotation of the maps was simultaneously applied to the quaternions data to keep the habit plane traces and the coherency of special boundary traces. Moreover, a resolution reduction by a factor of 2 was performed i.e. the information contained in a 2×2 pixels kernel was transformed into a single pixel by keeping the value of the top-left pixel. This artificially creates a dataset with a coarser step size and aims to provide robustness to the trained model against step size variations.

From each map, 11 additional maps have been generated (Figure 3a) and the training-validation set contained a total of 72 maps. All maps were cropped into 224×224 pixels frames without overlapping (Figure 3b). The crop size was chosen to contain enough information without much data loss and while being a multiple of $2^n$ (with $n$ the number of max. pooling operations in the U-Net; see section 3.4 ). The resulting crops were randomly shuffled and split with an 80-20 ratio: 864 crops to train and 216 crops to validate. Cropping and shuffling has been performed to show a wide variety of training examples and minimize overfitting. Additionally, it lowers the memory footprint during training.

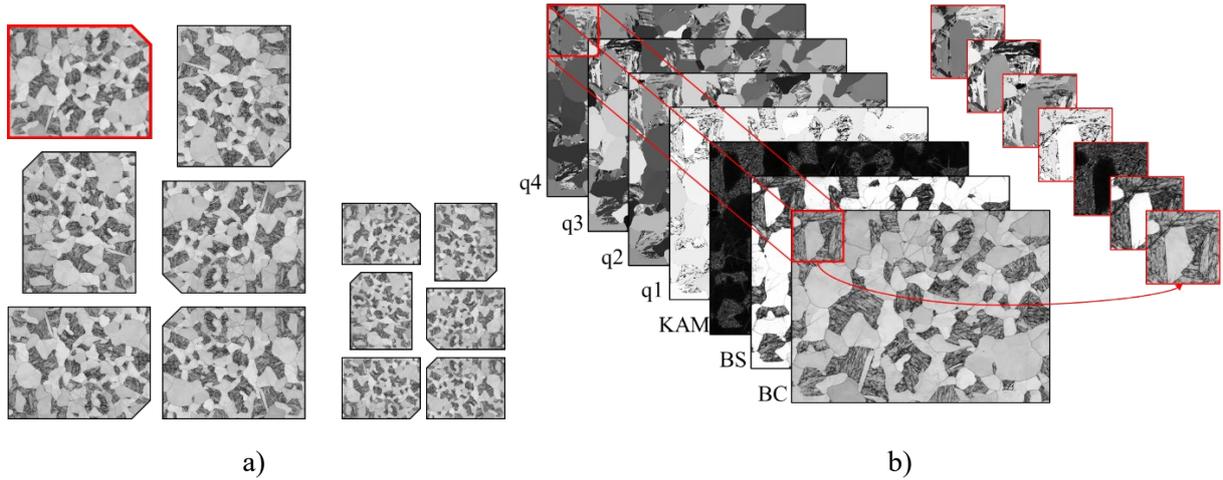

Figure 3. a) Data augmentation applied on the BC image of a particular EBSD map, providing 11 additional maps from the original one highlighted in red. The cut corner is just illustrative. b) A sliding window without overlapping is used along all the channels to obtain the training-validation crops.

## 3.4 Our U-Net and its hyperparameters

The implemented architecture is shown in Figure 4, it has a U-shape, with a contracting path, where the classical CNN steps convolution, activation and downsampling are applied, followed by a symmetric expansive path which consists in convolution, activation and upsampling. Additionally, skip connections enable the combination of high-resolution features from the contracting path with the expansive path to increase the resolution of the upsampled output.

Each 224×224 pixels crop contained in the training set enters the network to start the learning process, where 3×3 pixels convolution filters are applied with a padding "same" strategy. A "ReLu" activation function was chosen and the max. pooling/upsampling after each convolution step was done by a 2×2 pixels kernel. In total, the input data undergoes 7 convolutional steps and each one of them is formed by 2 convolution layers and a regularization layer in between, with 20% dropout. After the final 32-filters layer, a 1x1 pixel convolution is applied to map each 32-component feature vector to the desired number of classes. The final classification, where the probabilities are assigned to each of the candidate classes, is performed by a "Softmax" function. The model performance after each training iteration is evaluated on both the training and validation datasets by a "Categorical Cross Entropy" loss function. These last two were selected because the model will evolve to multi-phase classification in the future. Backpropagation learning is performed with the stochastic gradient descent method "adadelta". As mentioned in section 3.1 the network has been adapted to be able to receive multiple channels as input data enabling the training with several parameters at the same time.

The necessary code was implemented in Python using Tensor Flow [47] and Keras [48] libraries. All the models have been trained on a desktop computer equipped with an Intel Core i5-8250U CPU @ 1.60GHz, 16Gb of RAM memory without using the graphic card, taking ~6h per trained model. All the computation times expressed on this manuscript are relative to the aforementioned computer.

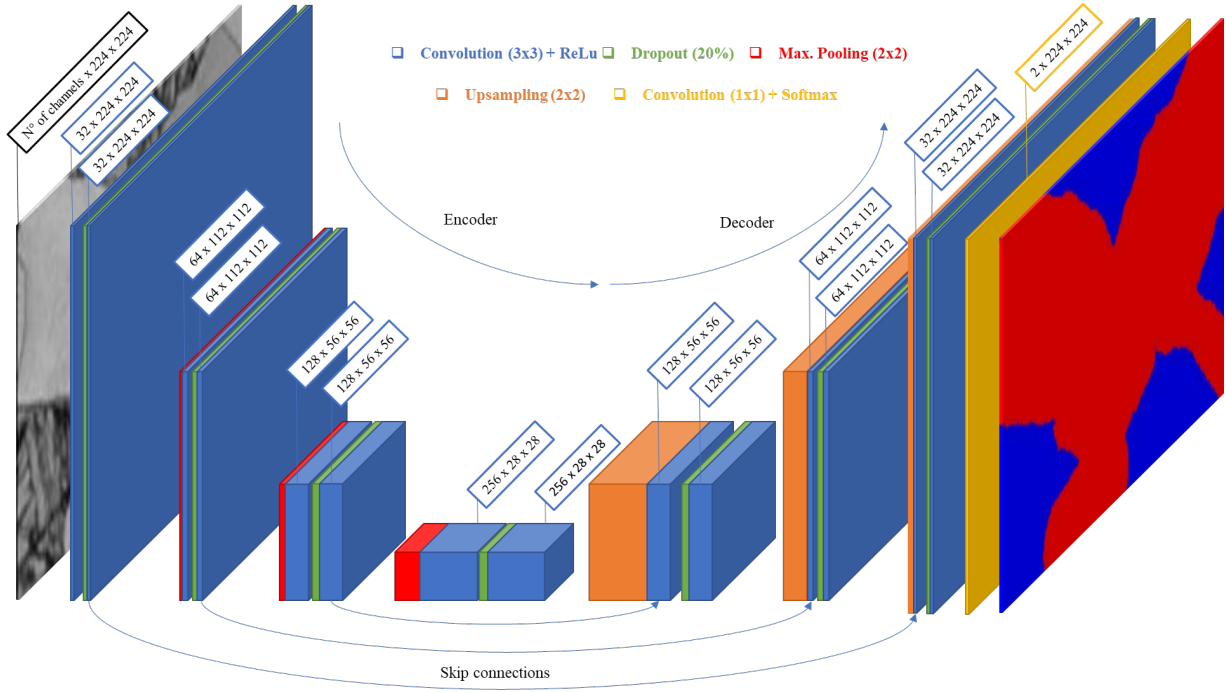

Figure 4. Scheme of the U-Net architecture used in this work.

### 3.5 Performance evaluation metrics

The performance has been calculated by comparing pixel-wise the ground truth labeling with the models' prediction. A prediction per pixel was considered correct when it matched the assigned label. The classification accuracy (CA) is described as the fraction of correctly predicted pixels among the whole number of pixels within the considered dataset. When required, a confusion matrix has been plotted to allow an in-depth visualization of the misclassifications. A confusion matrix is defined by:

- True Ferrite: pixels labeled as ferrite and predicted as ferrite by the model.
- True Martensite: pixels labeled as martensite and predicted as martensite by the model.
- False Ferrite: pixels labeled as martensite but predicted as ferrite by the model.
- False Martensite: pixels labeled as ferrite but predicted as martensite by the model.
- Precision: fraction of correctly predicted pixels of one class among all the predicted pixels of that class.
- Recall: fraction of correctly predicted pixels of one class among all the labeled pixels of that class.

## 4. Results

### 4.1 Performance on the training and validation datasets

Training and validation results are presented for all 4 models in Table 1, showing excellent CA above 98.5% for the BS, BC and KAM-based models. The model trained on quaternions achieved a very good CA overcoming 96.5%. In all the cases, the training and validation CA are almost identical which indicates that the models did not overfit the training dataset.

Table 1. Final training and validation CA reached by the 4 models.

|  | BS | BC | KAM | Quaternions |
|---|---|---|---|---|
| Training | 98.13 | 98.53 | 98.62 | 96.76 |
| Validation | 98.50 | 98.40 | 98.51 | 96.61 |

Figure 5 shows the training and validation CA against the number of epochs for the 4 models during the whole learning process. The number of epochs is equal to the number of times the network works on the entire dataset and optimizes its weights to improve the performance. High CA above 90% were achieved by all the models at early stages of the training process. At 15 epochs, both training and validation CA of the BC, BS and KAM-based models started to converge to the values presented in Table 1. At 25 epochs, the training CA of the quaternions-based model kept showing an upward trend while the validation CA remained constant, what indicates overfitting. For this reason, the model trained at epoch 25 was selected as the final one. Each epoch took ~700 s to be completed.

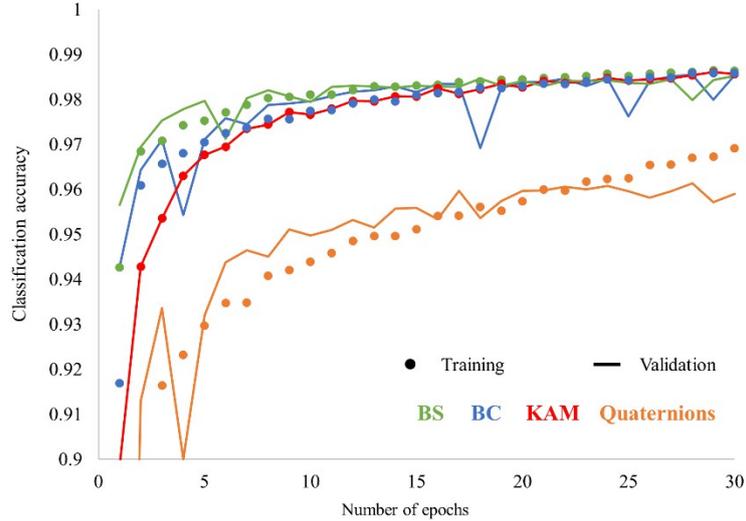

Figure 5. Classification accuracy of the 4 trained models plotted against the number of epochs.

### 4.2 Performance on the testing dataset

The performance was measured over the 9 maps of the testing dataset but cropped to 856×1136 pixels to support max. pooling/upsampling operations. Indeed, U-Net can be advantageously applied to any input size and it is not constrained by the input size used during training. This avoids the need to predict on several small frames and reconstruct the whole map afterwards, which induces a quality loss in the result. Regarding time, the classification result per map was obtained in less than 30 s once the desired input to work with was provided. The KAM and quaternions-based models took some additional preprocessing time in order to compute the KAM and homogenize symmetry equivalent for the quaternions.

The fractions of ferrite and martensite have been calculated with all 4 models and averaged over the 3 testing maps of each sample (Table 2). Compared to the ground truth labeling, the BS, BC and KAM-based models provide an error below 1% and the quaternions-based model a maximum error of 3%. This indicates an excellent agreement between the real and the predicted phase fractions.

Table 2. Fractions [%] of ferrite and martensite for each sample based on the average fractions within the 3 respective testing maps for all 4 models.

| Sample | | Labeling | BS | BC | KAM | Quaternions |
|---|---|---|---|---|---|---|
| T1 | Ferrite | 57.2 | 57.4 | 57.2 | 57.4 | 57.5 |
| | Martensite | 42.8 | 42.6 | 42.8 | 42.6 | 42.5 |
| T2 | Ferrite | 41.6 | 41.9 | 42.1 | 42.1 | 44.3 |
| | Martensite | 58.4 | 58.1 | 57.9 | 57.9 | 55.7 |
| T3 | Ferrite | 33.7 | 33.9 | 34.5 | 34.5 | 36.4 |

| | Martensite | 66.3 | 66.1 | 65.5 | 65.5 | 63.6 |

Pixel wise classification results are presented through the confusion matrices on Table 3. All the models present a high CA, meaning that the phase fractions calculated above are well funded. In addition, the obtained CA are very close to those of the training and validation datasets what highlights the robustness of the models since they can generalize to new examples. The greater part of false predicted pixels was found as false ferrite, indicating that for all the models it is more complex to correctly classify martensite than ferrite. This trend is even more pronounced for the quaternion-based model and can be also visualized in the lower values of ferrite precision and martensite recall (Table 3d).

Table 3. Confusion matrices of the a) BS-based model, b) BC-based model, c) KAM-based model and d) quaternions-based model over the testing dataset (>8.7 million pixels).

| | | Predicted class | | Recall |
|---|---|---|---|---|
| | | Ferrite | Martensite | |
| True class | Ferrite | 3792973 | 64466 | 98.32% |
| | Martensite | 114885 | 4779420 | 97.65% |
| Precision | | 97.06% | 98.66% | CA = 97.95% |

(a) BS-based

| | | Predicted class | | Recall |
|---|---|---|---|---|
| | | Ferrite | Martensite | |
| True class | Ferrite | 3800730 | 56709 | 98.52% |
| | Martensite | 133873 | 4760432 | 97.26% |
| Precision | | 96.59% | 98.82% | CA = 97.82% |

(b) BC-based

| | | Predicted class | | Recall |
|---|---|---|---|---|
| | | Ferrite | Martensite | |
| True class | Ferrite | 3787837 | 69602 | 98.19% |
| | Martensite | 89637 | 4804668 | 98.16% |
| Precision | | 97.68% | 98.58% | CA = 98.18% |

(c) KAM-based

| | | Predicted class | | Recall |
|---|---|---|---|---|
| | | Ferrite | Martensite | |
| True class | Ferrite | 3761912 | 95527 | 97.52% |
| | Martensite | 267297 | 4627008 | 94.53% |
| Precision | | 93.36% | 97.97% | CA = 95.85% |

(d) Quaternions-based

Figure 6 shows the classification results on 3 maps belonging to the testing set, representative of the different samples (from left to right: T1, T2 and T3 respectively). Figure 6a shows the BS images and the labeling, where red refers to ferrite and blue to martensite. Figure 6b presents the results for the BS-based model; the top row shows the image classified by the model while the bottom row shows the CA where pixels correctly classified are in white, false ferrite in blue and false martensite in red. Same results for the quaternions-based model appear in Figure 6c. Since the results obtained by the BS, BC and KAM are very similar we only present the BS-based model results.

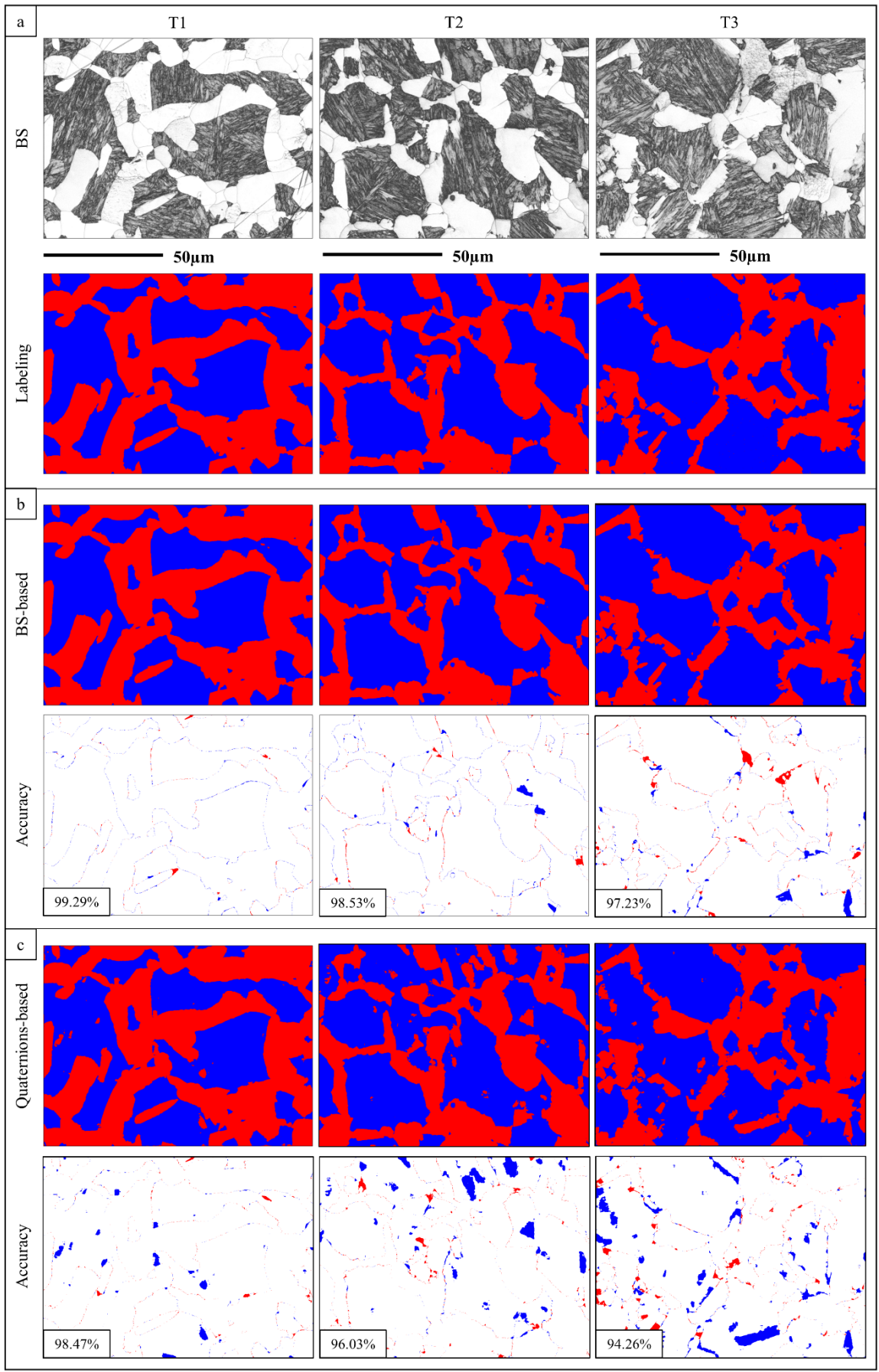

Figure 6. a) BS image and ground truth labeling of 3 maps belonging to the testing set (from left to right: T1, T2 and T3 respectively). Classification results and accuracy maps for the b) BS-based model and c) quaternions-based model.

T1 map has been classified with a CA above 99% using the BS-based model (Figure 6b.T1) and 98% for the quaternions-based model (Figure 6c.T1). As the complexity of the microstructure increases due to the increasing martensite fraction (Figure 6a.T2) and the presence of Widmanstätten ferrite (Figure 6a.T3), the CA drops to 98.53% (Figure 6b.T2) and 97.23% (Figure 6c.T2) for the BS-based model and to 96.03% (Figure 6b.T3) and 94.26% (Figure 6c.T3) for the quaternions-based model. Both models are robust to sample preparation artifacts as scratches (see in top-left and bottom-right corners of Figure 6a.T1) and surface roughness as in the ferritic grain at the top-center of Figure 6a.T3.

Three types of misclassifications have been identified. The first one is false ferrite located in large martensite laths with a low KAM value, high pattern quality indexes and convex shapes. These features likely correspond to auto-tempered martensite coarse lath and are relatively common in steels [49]. This is the largest source of error for all the models especially for the quaternions-based one as seen in Figure 6c where there is a larger fraction of false ferrite (blue clusters). The second type of misclassification concerns false martensite and correspond to small ferrite grains neighboring martensite domains. This small ferrite surrounded by martensite presents a lower BC and BS in average due to a high boundary-to-surface ratio. Indeed, EBSD patterns at the grain boundary have usually a lower quality due to pattern overlapping [50] and higher dislocations density [51]. The last source of errors is located at the interphase boundaries and is visible as thin colored layers in the accuracy map. The quaternions-based model is the less sensitive to this error followed by the KAM-based model. Since labeling was based on orientation data (grain detection) it is not surprising to have a better CA on boundaries detected with orientation derived inputs.

### 4.3 Stability of the method

The method's stability has been analyzed from 2 main perspectives: one inherent to the CNN learning process and the other one related to the potential errors introduced by the expert during labeling.

Regarding the first perspective, weight initialization, dropouts and gradient descent are stochastic processes and introduce some variability in the training. To evaluate this variability, 5 BS-based models have been trained from scratch presenting the training data exactly in the same order. The CA over the testing dataset presented a standard deviation of $\sigma = 0.075\%$ resulting in an uncertainty of $\pm 0.1\%$ (calculated as $3 \times \sigma/\sqrt{N}$ with N=5 being the number of trained models). The reproducibility of our models is then very good.

To evaluate the stability of the models against mislabeling, we have intentionally labeled 1% of the ferrite grains as martensite and 1% of martensite grains as ferrite for a total of 2% mislabeled data in the training set. The model trained in these conditions has obtained a testing CA of 97.3%, showing a 0.65% difference with the testing CA reached by the BS-based model (97.95%) presented in Table 3a. Although intentional mislabeling was added, the model remains robust and does not drastically reduce the CA. This highlights that errors that could be introduced during labeling by the expert (usually less than 2%) would only marginally affect the model's performances.

### 5. Discussion

### 5.1 Comparison to other state of the art approaches

The results obtained by our U-Net have been compared to 2 other approaches proposed in the literature to address the phase classification task. The first method relies on finding a manual threshold that separates ferrite from martensite based on grain averaged pattern quality values [11]. The second one is TWS since it has been proposed for optical micrographs segmentation [16], [26]. BS has been selected as the pattern quality parameter to work with these methods. They have been applied on the 3 maps presented in Figure 6.

The CA for all the methods on the 3 maps in Figure 6 is plotted in Figure 7. The best results on average over all 3 maps are achieved by our U-Net models (BS and KAM-based models with a CA of 98.35% and 98.48%, closely followed by our BC-based model with a CA of 97.81%). High CA (96.40% and 96.55%) are reached by both the grain averaged and TWS methods respectively. Our quaternions-based model presents a CA of 96.25%. Our BC, BS and KAM-based models overcome the rest of the approaches also in the complex microstructure of sample T3 (Figure 6). Our approach provides then slightly better results as the established state-of-the-art methods for phase discrimination.

Additionally, it must be noted that for both the grain average method and TWS, each map was treated individually what required time and effort to find the most suitable phase discrimination, while our approach classifies all the maps with the same trained model. TWS offers the possibility to train a model that can be saved to be applied on new maps without the need to relabel, but the model's loading can take several minutes depending on the trained model's size. In our approach, the models are instantly loaded and are able to classify a map in less than 30 s without the need of any user input.

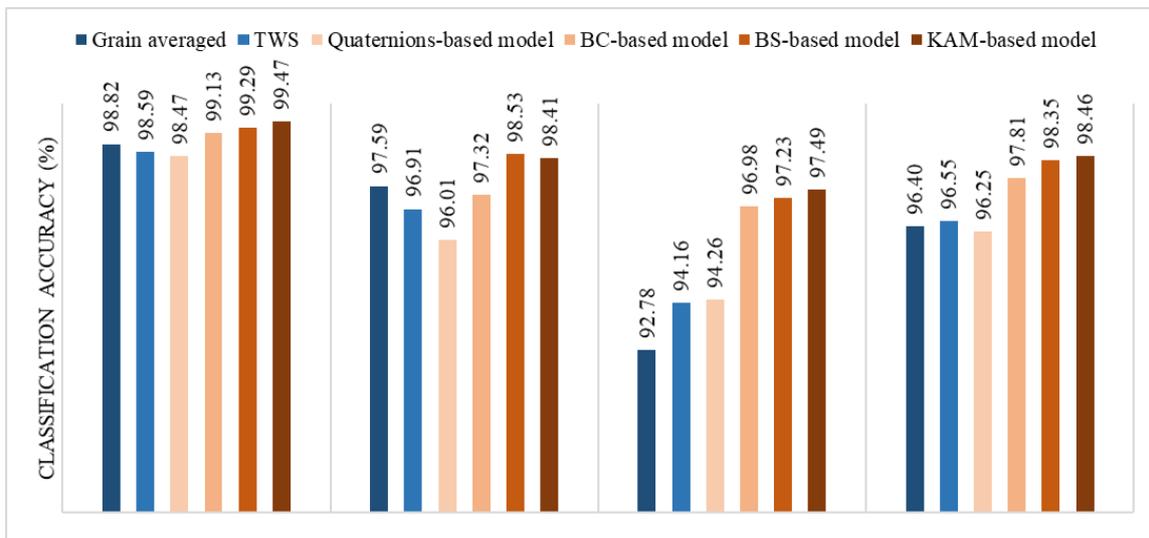

Figure 7. Classification accuracies reached by our 4 models and other state of the art methods over 3 representative microstructures of the testing set.

### 5.2 Models' optimizations

*U-Net architecture and input size*

Until now, "computation cost" has not been considered. The U-Net architecture presented in Figure 4 has been developed for high accuracies without caring about it. In this respect, several trials and errors have been carried out by reducing the network depth (7, 5 and 3 convolutional steps) and the input image size (224×224 and 100×100 pixels). Reducing the depth implies a reduction of the number of learning parameters whereas reducing the input image size means lower the memory footprint during training. For the 7 convolutional steps architecture, the 100×100 pixels images had to be enlarged to 104×104 pixels to be compliant with max. pooling operations.

Figure 8 plots the log of the testing error (1-CA) for the BS and quaternions-based models for the tested parameters. The input image size modification does not have a clear impact on the accuracy of the models. This means that the 100×100 pixels crops contain enough information to capture significant features to train the model, reducing the memory footprint and saving computation time. Training a 7-convolutional step architecture with an input image size of 104×104 pixels took ~120 s per epoch, compared to the ~700 s for the model presented in section 4.1.

Regarding network depth, the BS-based model is not substantially affected and keeps the testing error very low even with the shallower architecture. The situation is clearly different for the quaternions-based model, where significant improvements are observed when increasing the number of convolutional steps. This is due to the higher complexity of the input data, which requires to work with a higher number of learning parameters.

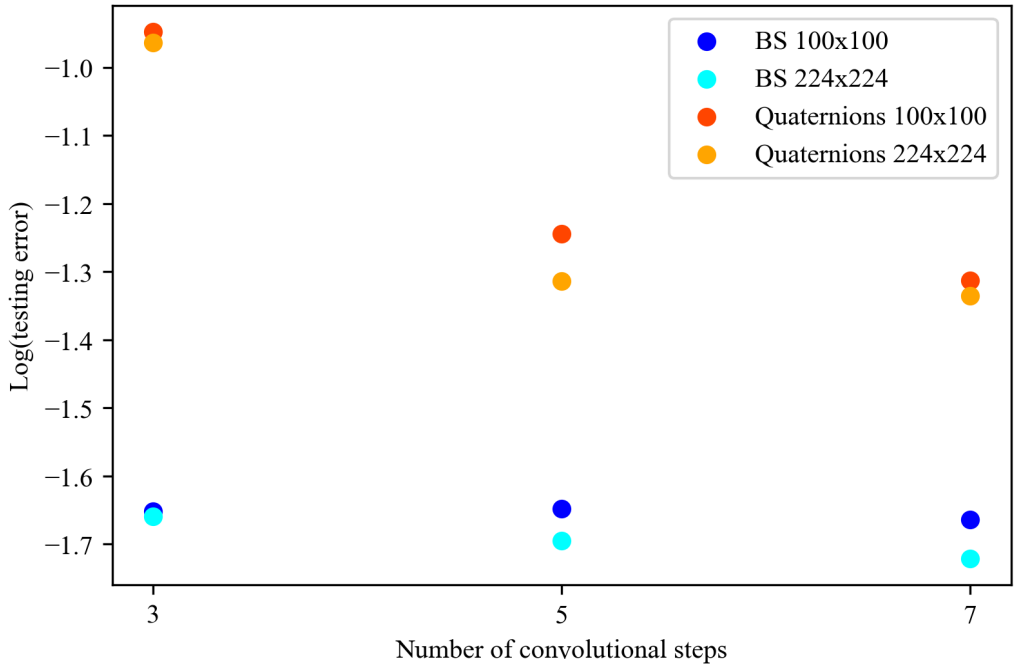

Figure 8. Effect of the number of convolutional steps and input image size for the BS and the quaternions-based models.

*Training set size and data augmentation*

Another important aspect to optimize is the amount of data required to train an efficient DL classifier. In the literature, it is reported that the error (1-CA) is an inverse power law of the amount of data up to a saturation point where an irreducible error always remains [52]. To assess this point, several models have been trained considering different amounts of training data with and without DA. An architecture with 5 convolutional steps and an input image size of 100×100 pixels has been selected to accelerate training.

Figure 9 plots the testing error for the quaternions and BS-based models trained with and without DA. The quaternions-based model benefits from increasing the number of crops in the training dataset. Applying DA is a good technique to improve the model's performance when the amount of available data is small. With this strategy, the power law function has a lower exponent, which means that data augmentation is not as efficient as providing never seen data. Once there is enough data, both curves converge to the mentioned saturation point.

For the BS-based model, the achieved performances are very high despite a small amount of data. This is a major asset when thinking about labeling, as it does not require a large number of training examples to achieve excellent results. As expected, DA provides a tiny improvement. In the last step, when considering more than 300 crops in the training set, there is a further improvement that may be related to capturing some specific new features that the model has not seen beforehand.

These results are coherent if we consider the complexity of the input data and how a human would proceed to perform the classification. As BS alone is a first order feature to discriminate ferrite and martensite by thresholding, not many training examples are needed. However, since orientation data is more complex to interpret, the amount of data needed increases.

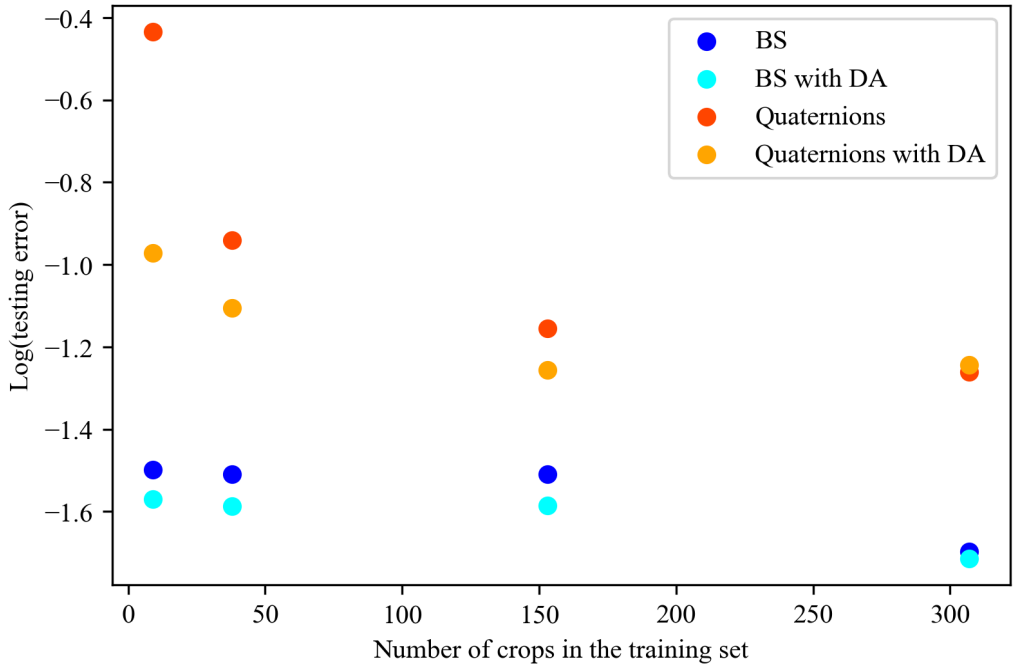

Figure 9. Effect of the training set size and data augmentation techniques for the BS and the quaternions-based models.

*Coupled inputs*

Since the input channels are different in nature, one can ask if coupling them would result in any improvement. As BS and BC had similar performances (accuracy and localization of errors), we ignored the BC in this assessment. This left 4 possible coupling to train new classifiers in the search of further improving the performance. The best results have been obtained with a BS+KAM-based model that overcome all the models presented above reaching a CA of 98.5% on the testing data. The confusion matrix in Table 4 explains that inputting these 2 parameters together improves the classification by reducing false ferrite at the expense of a slight increment in false martensite pixels. Figure 10a shows the BS image of a testing map belonging to T2 and Figure 10b its ground truth labeling. This map has been classified with BS-based model (Figure 10c), KAM-based model (Figure 10d) and BS+KAM-based model (Figure 10e). On this example, the BS+KAM has a CA of 99.15%. It takes advantages of the two previous models by reducing the false predicted ferrite and improving the interphase boundary classification. Having a closer look at the remaining errors in Figure 10e, the false ferrite clusters circled should have been labeled as ferrite. As stated, certain errors were introduced accidentally by the expert and in some cases the model could overcome them. Other couplings as BS+quaternions, KAM+quaternions and BS+KAM+quaternions have been tried without any significant improvement.

Table 4. Confusion matrix of the BS+KAM-based model over the testing dataset.

|  |  | Predicted class | | Recall |
|---|---|---|---|---|
|  |  | Ferrite | Martensite |  |
| True class | Ferrite | 3784863 | 72576 | 98.11% |
|  | Martensite | 58905 | 4835400 | 98.79% |
| Precision |  | 98.46% | 98.52% | CA = 98.5% |

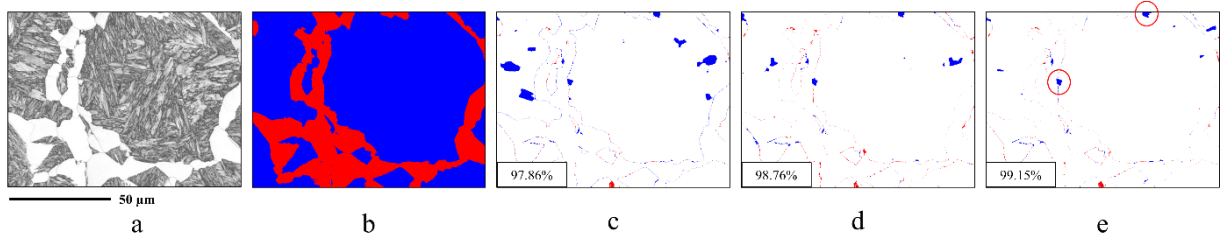

Figure 10. a) BS image of a T2 microstructure and b) its ground truth labeling. Accuracy maps obtained with c) BS-based model, d) KAM-based model and e) BS+KAM-based model.

### 5.3 Models' robustness

*Complex Dual Phase microstructure*

Figure 11a shows the BS image of a testing microstructure coming from a sample with the same composition detailed in 2.1 but that has undergone a different heat treatment: austenitization to 905°C followed by cooling in 5s to 640°C and a final quenching to room temperature. The map was obtained under the same conditions described in section 2.2. As a result, it can be observed a large fraction of martensite with complex lath shapes and different sizes, and in addition a great part of serrated ferrite/martensite interphases indicating a high fraction of Widmanstätten ferrite. This map looks rather different to what the model has been trained with and means an interesting challenge to classify. It has been labeled by an expert taking over 2 h to complete (Figure 11b). The best CA has been obtained with the BS+KAM-based model, overcoming 95% (Figure 11c). The main fraction of wrong predictions are false martensite pixels (red clusters in Figure 11d). They come mainly from the wrongly classified Widmanstätten ferrite plates, since the models have not been trained enough on this kind of configuration. Although the result is not perfect, it is a very good approximation to the ground truth labeling that can be obtained in much shorter times and with higher accuracies than the presented state-of-the-art approaches (91.6% with [11] and 85.4% with [34]).

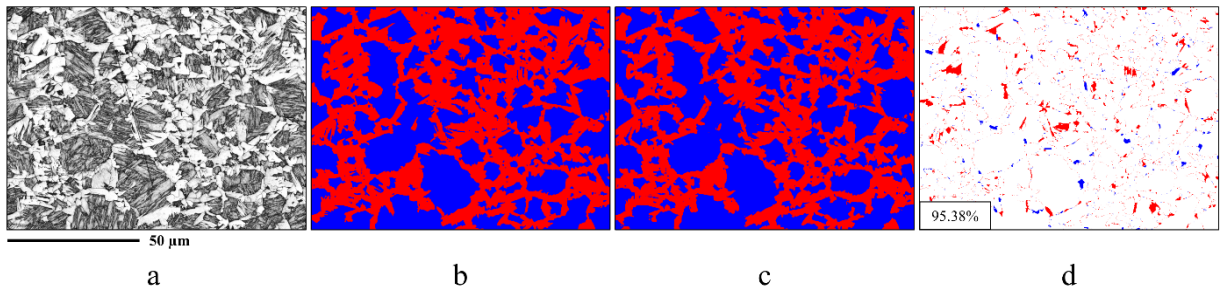

Figure 11. a) BS image and b) ground truth labeling of a microstructure that does not belong to the original dataset. c) BS+KAM-based model classification result and d) accuracy map.

### 5.4 Models' interpretation

Our results showed that most models easily learn that low BS, high KAM or domains with many non-indexed pixels correspond to martensite and reversely for ferrite. Nevertheless, it is not clear what kind of feature the quaternions-based model exactly extract from raw data as this task is more complex (even for a human). It can be noticed that the accuracies reached are lower than those obtained with the KAM, even if the KAM is a kind of convolution computed from raw orientation data.

To assess this, a reverse engineering process was applied by providing new data to the model: maps with averaged orientations per grain and maps where non-indexed pixels were cleaned. The orientation averaged map was classified with a high CA, meaning that the trained model does not look only at the small orientation variations within the grain to perform its classification. We can conclude that the network does not extract features similar to KAM from the quaternions alone. The result on the second map showed a very low CA, what means that the model uses the non-indexed pixels to carry out the classification. However, trained with mixed data (raw and corrected), the model could achieve similar results as previously. In the new model, recognized features seems to be the high density of grain boundaries, the lath-like shapes inside martensite and the grain sizes.

The U-Net architecture appear to be very well adapted to process images as BS, BC and KAM, but does not seem to be optimal for orientation data as provided in this work. This is likely due to the high complexity in feature extraction. For this reason, when inputs were coupled in section 5.2, no improvement was achieved with quaternions, since neural networks quickly reinforce the "easiest" way to perform the classification avoiding the complex feature extraction.

6. Conclusion

For the first time, a CNN method has been successfully implemented to classify steel microstructures from raw orientation data. A semantic segmentation approach using the U-Net architecture has been proposed.

Application was done on a DP steel where labeling could be done with a high accuracy. Although Widmanstätten ferrite almost systematically needed manual corrections.

High accuracies in the testing dataset have been obtained for all the models, reaching >95% for the quaternions-based and ~98% for the BS, BC and KAM-based models.

Compared to other state-of-the-art approaches for phase discrimination, the UNet model overcomes them in result accuracy and computation time.

Combining both BS and KAM as input further improved the classification. This model has been evaluated on microstructures with a that presented different features never seen during training and CA above 95% was reached.

A small dataset of 6 EBSD maps of 856×1141 pixels has been enough to create a representative training set. However, it has been demonstrated that the models' performance remains high even if the amount of available data is drastically reduced. Data augmentation techniques are helpful when starting from a very low amount of original data.

Several parameters of the model have been assessed:

- We showed that a shallower architecture of the U-Net was enough to capture first order features as BS, but a deeper one is needed for orientation data.

- Intentional errors introduced during labeling showed that the model was robust to mislabeling.

As the U-Net architecture is not well adapted to work with raw orientation data, the challenge for the coming studies on multiphase steel microstructures is on finding a proper architecture or a better way to provide the orientation data in order to exploit its full potential.

Acknowledgements

This work was supported by the French State through the program "Investment in the future" operated by the National Research Agency (ANR) and referenced by ANR-11-LABX-0008-01 (Laboratory of Excellence 'DAMAS': Design of Alloy Metals for low-mAss Structures).

LG and NG acknowledge advises of Hervé Frezza-Buet and Jeremy Fix, from Loria, Supelec about Machine Learning.

Data availability

The raw/processed data required to reproduce these findings are available on direct demand to the authors.